\documentclass[twocolumn,showpacs,preprintnumbers,amsmath,amssymb]{revtex4}

\usepackage{graphicx}

\usepackage{dcolumn}
\usepackage{bm}
\usepackage{verbatim} 
\usepackage{units}
\usepackage{tipa}

\newcommand{\be}{\begin{equation}}
\newcommand{\ee}{\end{equation}}
\newcommand{\bea}{\begin{eqnarray}}
\newcommand{\eea}{\end{eqnarray}}

\newcommand{\LiS}{^{6}\mathrm{Li}}

\newcommand{\gl}{\gamma_{\mathrm{loss}}} 

\newcommand{\Nfty}{N_{\infty}}

\newcommand{\NLis}{N_{\infty}}

\newcommand{\GLi}{\Gamma}

\newcommand{\dnp}{\Delta\nu_{\mathrm{p}}}
\newcommand{\dnr}{\Delta\nu_{\mathrm{r}}}

\newcommand{\Ir}{I_{\mathrm{r}}}
\newcommand{\Ip}{I_{\mathrm{p}}}

\newcommand{\maxli}{(8 \pm 2) \times 10^7}

\newcommand{\maxlir}{(1.3 \pm 0.3) \times 10^6 ~\mathrm{s}^{-1}}

\newcommand{\aB}{a_{\mathrm{B}}}

\newcommand{\one}{|1\rangle}
\newcommand{\two}{|2\rangle}

\newcommand{\ut}{U_{\mathrm{trap}}}

\newcommand{\kb}{k_{\mathrm{B}}}

\newcommand{\btof}{690~\mathrm{G}}

\newcommand{\td}{T^*}

\newcommand{\vb}{\bar{v}}
\newcommand{\nubar}{\bar{\nu}}
\newcommand{\gp}{\gamma_{\mathrm{peak}}}
\newcommand{\zr}{z_{\mathrm{R}}}

\newcommand{\nmot}{N_{\mathrm{MOT}}}
\newcommand{\nodt}{N_{\mathrm{ODT}}}
\newcommand{\todt}{T_{\mathrm{ODT}}}
\newcommand{\uodt}{U_{\mathrm{ODT}}}

\newcommand{\tevap}{\tau_\mathrm{{evap}}}
\newcommand{\tc}{T_\mathrm{{c}}}
\newcommand{\tf}{T_\mathrm{{F}}}

\newcommand{\frpos}{832.18~\mathrm{G}}
\newcommand{\fr}{\mathrm{FR}}

\newcommand{\lodt}{\lambda_{\mathrm{ODT}}}
\newcommand{\cdt}{\mathrm{CODT}}

\newcommand{\amfg}{\partial_z|B|}

\newcommand{\amol}{a_{\mathrm{mol}}}

\begin{document}

\date{\today}
\title{Realization of BEC-BCS crossover physics in a compact oven-loaded magneto-optic trap apparatus}

\author{Will Gunton, Mariusz Semczuk, and Kirk W.~Madison}

\affiliation{Department of Physics and Astronomy, University of British Columbia, Vancouver, Canada
}

\begin{abstract}
We report on a simple oven-loaded magneto-optical trap (MOT) apparatus for the creation of both molecular Bose-Einstein condensates (mBEC) and degenerate Fermi gases (DFGs) of lithium.  The apparatus does not require a Zeeman slower or a 2D MOT nor does it require any separation or differential pumping between the effusive atom source and the science chamber.  The result is an exceedingly simple, inexpensive, and compact vacuum system ideal for miniaturization.  We discuss our work in the context of other realizations of quantum degenerate gases by evaporation in optical dipole traps and illustrate that our apparatus meets the key requirements of atom number and trap lifetime.  With this system, we also demonstrate the production of a mBEC, and we use it to observe the pairing gap of a strongly interacting two-component DFG in the BEC-BCS crossover regime.
\end{abstract}


\maketitle

\section{Introduction}

Since the first achievement of Bose-Einstein condensation (BEC) in atomic gases, techniques for the production of quantum degenerate gases (QDGs) from laser cooled atoms have been continually improved, refined and simplified.  Because the phase space density achievable with laser cooling is limited by light-induced, density-dependent collisional losses and by the temperature limits of laser cooling, QDGs are typically made in a two step process.  The only exception to this is Sr which has been directly laser cooled to degeneracy \cite{PhysRevLett.110.263003}.  After laser cooling, the ensemble is transferred into a conservative potential formed by an off-resonant magnetic or optical field and forced evaporative cooling is performed to achieve the requisite temperatures and densities.  Efficient evaporative cooling relies on rapid thermalization of the atoms in the trap and requires that the evaporation time be shorter than the lifetime of a particle in the trap.  Since the speed of evaporation is limited by the elastic collision rate, $\gamma = n \sigma \vb$ (where $n$ is the atomic density, $\sigma$ is the elastic collision cross section, and $\vb$ is the average collision velocity), this requirement places simultaneous constraints on the minimum initial atom number loaded into the conservative trap and on the quality of the vacuum system because the density of room-temperature, background-gas particles sets an upper bound on the trap lifetime.  These two constraints are in conflict since the atom number available for the initial stage of laser cooling is proportional to the partial pressure of those atoms, and it is the presence of this atomic vapor that also ultimately limits the trap lifetime.  In this work, we demonstrate that these requirements can be met for the production of QDGs of lithium by a magneto-optic trap loaded from a simple effusive atomic source.  This apparatus does not require a separated cold atom source such as a Zeeman slower \cite{PhysRevLett.48.596,miranda99} or a 2D MOT \cite{PhysRevA.80.013409} or an additional 3D MOT from which the atoms are transferred \cite{faucet}.  Nor does it require any separation or differential pumping between the effusive atom source and the science chamber.  The result is a small and inexpensive vacuum system ideal for miniaturization.  We fully characterize our system and demonstrate with it the production of a mBEC, and we use it to observe evidence of the pairing gap in a strongly interacting two-component DFG in the BEC-BCS crossover regime.

\section{Background}

\begin{table}[h]
\renewcommand{\arraystretch}{1.0}
\centering
\caption[c]{Reported experimental parameters for QDG production using ODTs loaded directly from a MOT.
$\nodt$ is the total atom number including all spin states and corresponds to twice the number of Feshbach molecules for those experiments that report on a molecular BEC.
}
\label{table:MOTandODTParams}
\begin{tabular}{c|ccc|c|ccc|c}
  \hline
  \hline
$^{87}$Rb & \multicolumn{3}{c|}{initial conditions} & Evap. & \multicolumn{3}{c|}{final conditions} & Ref. \\
 $\lodt$  & $\nmot$ & $\nodt$ & $\uodt$ & $\tevap$ & $\nodt$ & $\todt$ & $T/\tc$ & \\
$(\mu$m) & ($10^8$) & ($10^6$) & (mK) & (s) & ($10^5$) & (nK)  &  & \\
  \hline
10.6 & $0.3$ & 2 & 0.13\footnote{This quantity was inferred from other information.} & 2 & 1.8 & 375  & 1 & \cite{PhysRevLett.87.010404} \\

  10.6 & 10 & 4 & 1.7$^a$ & 5 & 3 & 160  & 1 & \cite{Gericke07}\\

1.96 & 5 & 2 & 0.19 & 20 & 0.1 & $<70$  & $\ll 1$ & \cite{PhysRevA.83.035601}\\

1.07 & $0.4$ & 0.35 & 0.6 & 12 & 0.4 & 140  & 1 & \cite{PhysRevA.84.043641}\\

1.064  & $3$ & 1.7 & 2 & 3 & 2 & 250  & 1 & \cite{Arnold20113288} \\

  \hline
1.565  & 30 & 3 & 0.65$^a$ & 3.5 & 3 & -- & 1 & \cite{PhysRevA.79.061406}\footnote{the atoms were transferred into the ODT from a large volume ODT loaded from the MOT} \\

1.55  & $\sim 1$ & 0.5 & 0.60$^a$ & 4.35 & 0.2 & -- & $\ll1$ & \cite{PhysRevA.87.053613}$^{b}$\\

  \hline
  \hline

$^{23}$Na & \multicolumn{3}{c|}{initial conditions} & Evap. & \multicolumn{3}{c|}{final conditions} & Ref. \\
 $\lodt$ & $\nmot$ & $\nodt$ & $\uodt$ & $\tevap$ & $\nodt$ & $\todt$ & $T/\tc$ & \\
$(\mu$m) & ($10^8$) & ($10^6$) & (mK) & (s) & ($10^5$) & (nK)  &  & \\
  \hline
1.07 & 0.1 & 0.3 & 1$^a$ & 2 & 0.2 & 1000  & $1$ &  \cite{1367-2630-13-6-065022}\\
  \hline   
  \hline
$^{133}$Cs & \multicolumn{3}{c|}{initial conditions} & Evap. & \multicolumn{3}{c|}{final conditions} & Ref. \\
 $\lodt$ & $\nmot$ & $\nodt$ & $\uodt$ & $\tevap$ & $\nodt$ & $\todt$ & $T/\tc$ & \\
$(\mu$m) & ($10^8$) & ($10^6$) & (mK) & (s) & ($10^5$) & (nK)  &  & \\
  \hline
1.064 & -- & 2 & 0.003 & 4 & 5 & 64  & $<1$ &  \cite{PhysRevA.78.011604} \\
  \hline
  \hline
$^7$Li & \multicolumn{3}{c|}{initial conditions} & Evap. & \multicolumn{3}{c|}{final conditions} & Ref. \\
 $\lodt$ & $\nmot$ & $\nodt$ & $\uodt$ & $\tevap$ & $\nodt$ & $\todt$ & $T/\tc$ & \\
$(\mu$m) & ($10^8$) & ($10^6$) & (mK) & (s) & ($10^5$) & (nK)  &  & \\
  \hline
1.07 & $10$ & 1 & 2 & 3 & 0.06 & 380  & $1$ &  \cite{PhysRevA.77.023604} \\
  \hline
  \hline
 $^6$Li & \multicolumn{3}{c|}{initial conditions} & Evap. & \multicolumn{3}{c|}{final conditions} & Ref. \\
$\lodt$ & $\nmot$ & $\nodt$ & $\uodt$ & $\tevap$ & $\nodt$ & $\todt$ & $T/\tf$ & \\
$(\mu$m) & ($10^8$) & ($10^6$) & (mK) & (s) & ($10^5$) & (nK)  &  & \\
  \hline
10.6 & -- & 2 & 0.55 & $\sim2.5$ & 3 & 360  & 0.55 & \cite{1367-2630-8-9-213}\\

1.09 & $0.2$ & 1.3 & 3.5 & 4 & 0.4 & 200 & $0.2$ & $^{\mathrm{this}}_{\mathrm{work}}$\\

1.07 & $10$ & 0.2 & 3 & 7.5 & 2.5 & 50  & $<0.2$ &   \cite{PhysRevA.81.043637}\footnote{Here $^6$Li was a coolant for $^{40}$K}\\

1.064 & $\sim 1$ & 5 & 2 & 3.6 & 3.6 & 1900$^a$  & 0.5 & \cite{PhysRevLett.102.165302}\\

1.064 & $\sim 1$ & 5 & 2 & 3.6 & 1.8 & 50  & 0.28 & \cite{PhysRevLett.103.130404}\\

1.064 & $\sim 1$ & 0.4 & 1.4 & 3 & 0.2 & 250$^a$ & $<0.25$ &  \cite{0953-4075-40-20-011} \\

  \hline  
1.064 & 10 & 1.5 & 0.15 & -- & 6 & -- & $0.3$ &  \cite{1367-2630-13-4-043007}\footnote{atoms were transferred into the ODT from a large volume resonator ODT loaded from the MOT}\\
1.03 & -- & 1.5 & 0.8$^a$ & 2 & 3 & 50 & $<0.2$ &  \cite{Jochim19122003}$^d$ \\

  \hline
1.070 & 15 & 6 & 0.56 & 6 & 3 & 1900 & $0.45$ &  \cite{PhysRevA.84.061406}\footnote{atoms were transferred into the ODT from a narrow-line laser MOT}\\
  \hline
\end{tabular}
\end{table}

In many early experiments with QDGs, evaporative cooling was performed in macroscopic magnetic traps where, due to the weak confinement, an initial ensemble size of $10^8$ to $10^9$ atoms required an evaporation time of 30 seconds or more.  To obtain this large trapped atom number and a long trap lifetime, a high flux, pre-cooled atomic beam generated in a region of high vapor pressure is recaptured by a magneto-optic trap in a region of low vapor pressure \cite{PhysRevLett.48.596,miranda99,PhysRevA.80.013409, Myatt:96, faucet}.  This approach allows the atomic source to be isolated by differential pumping from the region of evaporation.  With the development of smaller, sub-mm sized magnetic traps, the trapping confinement could be enhanced and the initial atom number required for evaporation was consequently reduced allowing for the production of a BEC in the same vapor cell as the MOT \cite{Hansel04102001}.  For a harmonic potential, the peak collision rate for a Boltzmann gas at the center of the trap is $\gp \propto N \sigma \nubar^3 / T$, where $N$ is the total atom number, $\nubar = (\nu_x \nu_y \nu_z)^{1/3}$ is the geometric mean of the trap frequencies, and $T$ is the ensemble temperature.  Thus, a small initial particle number can be offset by a stronger trap confinement or a lower initial temperature.  The production of a quantum degenerate Fermi gas of $^{40}$K by sympathetic cooling with $^{87}$Rb was also demonstrated with a chip based micro-scale magnetic trap loaded from an atomic vapor \cite{Thywissen-chip-nature}.  In that experiment, the evaporation time was 6 seconds and the the MOT loading of Rb was enhanced while keeping the background pressure low by light-induced atom desorption resulting in a trap lifetime of 9 seconds \cite{chipqdg}.

The use of high-power optical dipole traps (ODTs) to generate the conservative potential has several advantages over a magnetic trap including the ability to trap multiple spin states.  Indeed, the evaporation of a spin polarized Fermi gas in a magnetic trap is blocked at low temperatures because $s$-wave collisions are forbidden by the exchange statistics and the cross section for higher partial wave collisions rapidly drops to zero with temperature (the $p$-wave cross section falls to zero as $T^2$).  An important limitation of performing forced evaporation from an ODT by simply reducing the optical power is that this lowers both the trap depth and the trap frequencies leading to a continual decrease in the collision rate as evaporation progresses \cite{PhysRevA.64.051403}.  This is in contrast to forced evaporation in magnetic traps where the depth can be lowered by the use of a radio-frequency ``knife" without changing the confinement.  This latter method results in runaway (accelerating) evaporation when the ensemble temperature falls faster than the atom number.  For evaporation from an ODT, the decreasing collision rate will result in an eventual stagnation of the evaporation process and a limit to the maximum gain in phase space density. Nevertheless, this simple evaporation technique has been made to work for the production of a BEC of $^{87}$Rb atoms starting from a vapor loaded MOT transferred into an optical dipole trap formed with $\lambda = 10$~$\mu$m light.  The initial number in the ODT was $2 \times 10^6$ atoms, the trap lifetime was 6 seconds, and the evaporation time was 2.5 seconds \cite{PhysRevLett.87.010404}.

An additional problem associated with evaporation in a dipole trap is that for a fixed trap size (fixed by the minimum beam width, $W_0$) the use of a shorter wavelength laser results in a longer Rayleigh length ($\zr = \pi W_0^2/\lambda$) due to the slower beam divergence.  Since the Rayleigh length determines the longitudinal confinement, the slower beam divergence associated with the shorter wavelength results in a weaker confinement and less efficient evaporation.  This problem persists for a crossed dipole trap configuration since the slower beam divergence will allow a non-negligible fraction of atoms to reside far outside of the crossing leading to the same inefficiencies in evaporation.  A recent paper discusses these issues and presents a simple evaporation scheme to produce a BEC of $^{87}$Rb atoms from a vapor cell loaded MOT using a crossed ODT ($\cdt$) produced with light from a multi-longitudinal-mode laser operating at $\lambda = 1070$~nm \cite{Arnold20113288}.
In that work, the vacuum lifetime was 6 seconds and the evaporation time was 3 seconds.  Adding a mobile lens system to increase the confinement during evaporation has been shown in samples of  $^{87}$Rb to provide even more efficient trap loading and evaporation resulting in larger BEC ensembles \cite{PhysRevA.71.011602}.  Alternate strategies to perform more efficient evaporation without relaxing the optical trap confinement using either a strong magnetic field gradient \cite{PhysRevA.78.011604} or a displaced auxiliary ODT beam have been demonstrated \cite{PhysRevA.79.061406}.  It should be noted that for the case of a resonantly interacting $^6$Li gas, where the collision cross section is unitarity limited by the de Broglie wavelength, runaway evaporation in a simple ODT has been shown possible \cite{1367-2630-8-9-213}.

\begin{figure}[ht]
  \begin{center}
    \includegraphics[width=0.49\textwidth]{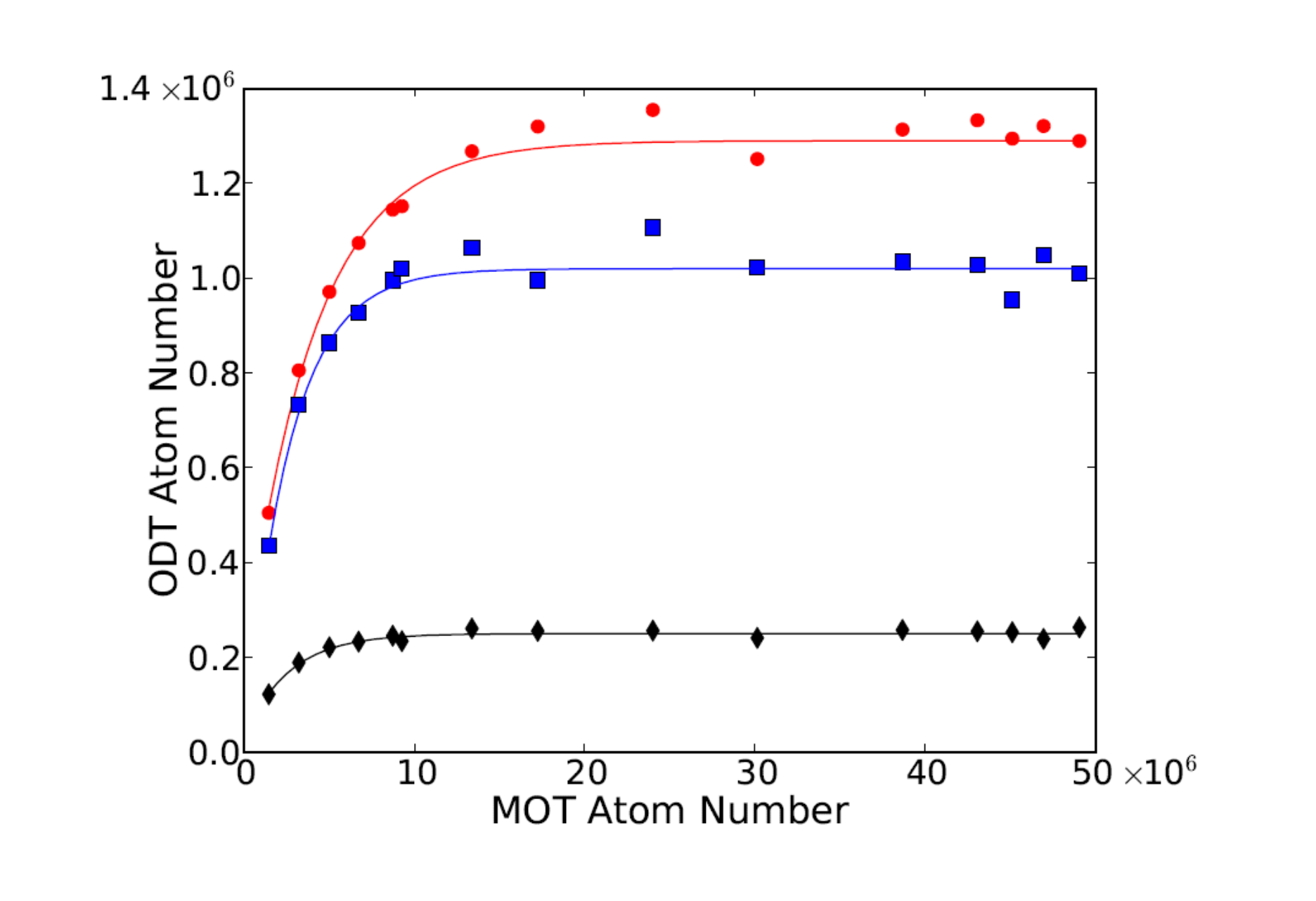}
\caption{(color online) Atom number transferred into the $\cdt$ versus atom number in the MOT.  Plotted are the initial atom number in the high power $\cdt$ at 200 W total power (circles), the atom number in the high power $\cdt$ after evaporation to 100 W total with 50 W per beam (squares), and the atom number transferred into the low power $\cdt$ (diamonds) for various initial MOT atom numbers.}
\label{fig:CODTloading}
  \end{center}
\end{figure}

An important observation is that for experiments reporting on the creation of a QDG in an ODT loaded from a ``standard" MOT (operating on the D2 line), the initial atom number transferred into the ODT is almost without exception between $4 \times 10^5$ and $5 \times 10^6$ atoms, and these numbers are achieved with initial MOT numbers in the range from $3 \times 10^7$ to $10^8$ atoms (see Table~\ref{table:MOTandODTParams}).  Even for experiments where the initial MOT number could be made as high as $10^9$ atoms by prolonging its loading, this is not typically done since the additional MOT atom number does not significantly increase the ODT number (e.g.~see \cite{PhysRevLett.102.165302}).  An example of this atom number saturation during transfer can be seen in Fig.~\ref{fig:CODTloading} where for our ODT geometry the number in the dipole trap does not increase substantially for MOT numbers larger than $2 \times 10^7$.  This saturation is, in large part, the result of density dependent losses that occur during the loading of the ODT from the MOT.  These are the same losses that limit the density of atoms in the MOT; however, they are typically more severe in the ODT because it has a much lower trap depth (in the milikelvin or microkelvin range) than the MOT (typically in the kelvin range).  The density dependent losses in the ODT include spin exchange collisions, hyperfine changing collisions, and light-assisted collision loss processes including photoassociation, fine-structure changing collisions, and so-called radiative escape (RE) (a detailed review of cold collisions can be found in Ref.~\cite{RevModPhys.71.1}).  RE is the dominant loss mechanism in a MOT and occurs when an atom pair absorbs a single photon during a collision leading to a large attractive dipole-induced interaction between the atoms for a brief period of time before the photon is reemitted.  This attraction accelerates the atoms and leads to a large increase of their kinetic energy such that they can escape the trap.  Even when considerable effort is made to mitigate density dependent collisional losses, still only $8 \times 10^6$ $^{87}$Rb atoms could be loaded into an ODT \cite{PhysRevA.62.013406}.  The dynamics of the ODT loading and loss mechanisms have been studied in detail previously \cite{PhysRevA.62.013406,PhysRevA.63.043403,Sofikitis,PhysRevA.79.013418}, and for sufficiently deep ODTs (whose depth is much larger than the temperature of atoms in the MOT reservoir), the atom number loaded into them from the MOT is primarily related to the ODT volume.  An additional complication is that the deeper the trap is made the more slowly it loads and reaches equilibrium \cite{PhysRevA.63.043403}.

An approach to achieving a deep and large volume ODT with limited optical power available is to use the enhancement provided by a passive optical resonator \cite{Jochim19122003,1367-2630-13-4-043007}.  The disadvantage is that while these large volume ODTs capture many more atoms than a smaller volume ODT, they do not have large trapping frequencies and therefore the atoms must be transferred into a smaller volume ODT where rapid forced evaporation is possible.  In Ref.~\cite{1367-2630-13-4-043007} more than $9 \times 10^7$ $^6$Li atoms were loaded from a MOT of $10^9$ atoms into a large volume resonator trap.  From this ensemble, $1.5 \times 10^6$ atoms at a temperature of $T=15$~$\mu$K were then transferred into a tightly focused ODT.  While the number loaded into the small volume ODT is smaller than that achieved by directly loading from the MOT, the temperature is an order of magnitude smaller providing an initial phase space density that is several hundreds of times larger.  The result is that the evaporation need not be as deep and the final ensemble size after crossing the degeneracy threshold is substantially larger than without this intermediate loading step (see Table~\ref{table:MOTandODTParams}).  In the case of Rb, strong sub-doppler cooling can provide substantially colder ensembles than in Li and a large volume ODT can be formed without passive enhancement provided by a cavity.  In addition, when the ODT is formed from 1565~nm light, the differential ac Stark shift for the laser cooling transition at 780~nm in Rb is very strong, and the ODT forms a so-called ``dark MOT" in the region of overlap.  This suppresses the light assisted collisional losses and enhances the number that can be transferred from the MOT to the ODT.  Ref.~\cite{PhysRevA.79.061406} reports loading $3 \times 10^7$ atoms from a MOT of $3 \times 10^9$ $^{87}$Rb atoms into such a large volume ODT at 1565~nm.  From this trapped gas, $3 \times 10^6$ atoms at a temperature of $T=65$~$\mu$K were then transferred into a tightly focused ODT, and a 3.5 s evaporation stage produced an ensemble at the BEC critical temperature with $3 \times 10^5$ atoms.  A similar experiment transferred $5 \times 10^5$  $^{87}$Rb atoms at a temperature of $T=60$~$\mu$K into a small volume ODT from a large volume ODT formed from 1550~nm.  From these initial conditions, a pure BEC of $2 \times 10^4$ atoms was formed after 4.35~s of evaporation \cite{PhysRevA.87.053613}.  It should be noted that the ac Stark shift induced by an ODT operating at wavelength above 1530~nm moves the laser cooling transition in Rb to smaller frequencies leading to an apparent blue shift of the laser cooling light for atoms inside the ODT.  While this blue shift is more effective than a red shift for the suppression of the absorption of repump light \footnote{The repump light for a $^{87}$Rb MOT is usually operated on resonance with the field free $F=1 \rightarrow F'=2$ transition, and, in the case of a red shift induced by the ODT, it would be shifted closer to resonance with lower frequency transitions to lower ($F'=1$ or $F'=0$) hyperfine excited states which do not decay back to the $F=2$ ground state producing not a suppression but rather an enhancement of the absorption of repump light leading to a ``bright MOT" inside the ODT.} and the realization of a ``dark MOT" in the ODT, it necessitates the MOT laser cooling light to be strongly red detuned during transfer to the ODT so that the cooling transition for atoms inside the ODT is not shifted below the incident light's frequency leading to Doppler heating and ejection of atoms from the ODT.

A related two-step loading strategy is to transfer the atoms from the MOT into a large-volume magnetic trap, perform some preliminary evaporative cooling, and then transfer the remaining atoms into a small volume ODT.  The disadvantage is that the preliminary evaporation stage in the magnetic trap necessitates a large initial ensemble number and a high quality vacuum.  The details of such hybrid trap strategies have been discussed in \cite{PhysRevA.73.043410}, and the production of $^{87}\text{R}\text{b}$ Bose-Einstein condensates in a combined magnetic and optical potential has been demonstrated \cite{PhysRevA.79.063631}.  This latter demonstration resulted in a condensate of $2 \times 10^6$ $^{87}\text{R}\text{b}$ atoms.  This strategy has also been used to transfer up to $3 \times 10^7$ $^{39}$K atoms into a small volume ODT from a magnetic trap and allowed for the production of  $^{39}$K condensates of $8 \times 10^5$ atoms \cite{PhysRevA.86.033421}.  This achievement is significant because the evaporation of $^{39}$K to degeneracy in a magnetic trap is particularly difficult given the Ramsauer minimum in the collisional cross section at 400~$\mu$K.

Another related two-step approach to achieving a larger QDG after the final evaporation demonstrated for $^6$Li is to transfer the atoms from the standard MOT (operating on the D2 line) into a MOT operating on a much more narrow transition.  For lithium, where sub-Doppler cooling is inhibited in the MOT (because the excited state hyperfine splitting is unresolved), recapturing the atoms in an ultraviolet MOT operating on the 
$2S_{1/2} \rightarrow 2P_{3/2}$ transition produces a much colder ensemble (by almost a factor of ten) that can then be loaded into a more shallow ODT.  This results in an ensemble with a much higher initial phase space density that can be cooled to degeneracy with less particle loss.  In Ref.~\cite{PhysRevA.84.061406}, $6 \times 10^6$ atoms were loaded into a $\cdt$ at $T=60$~$\mu$K and after 6 seconds of evaporation a DFG was produced with $3 \times 10^6$ atoms at $T=1.9$~$\mu$K ($T/\tf=0.45$).

In summary, achieving quantum degeneracy by rapid evaporation in a tightly confining ODT has been shown possible with initial numbers as low as $3 \times 10^5$ atoms \cite{1367-2630-13-6-065022}, and sufficient ODT numbers for most species has been achieved starting with MOT populations in the range from $10^7$ to $10^8$ atoms (see Table~\ref{table:MOTandODTParams}).  Furthermore, because of density dependent losses, the ODT number (for typical trap volumes surveyed in Table~\ref{table:MOTandODTParams}) saturates quickly for MOT sizes above $10^8$ atoms.  The consequence is that empirical evidence suggests that producing a QDG in an ODT is possible if on the order of $10^8$ atoms can be loaded in a MOT with a trap lifetime limited by background collisions of 10 s or longer.

In this work, we demonstrate that these conditions can be met for $^6$Li with a simple oven-loaded MOT.  Our apparatus does not require a Zeeman slower or a 2D MOT nor does it require any separation or differential pumping between the effusive atom source and the science chamber.  The result is an exceedingly simple vacuum system ideal for miniaturization.  We transfer $1.3 \times 10^6$ $^6$Li atoms into a $\cdt$ from a MOT of $2 \times 10^7$ atoms accumulated from the atomic beam flux in 20~s.  With this system, we demonstrate the production of a mBEC, and we show that it can be used to measure the pairing gap of a strongly interacting two-component DFG in the BEC-BCS crossover regime.  The key results of this work are that (1) a sufficient number of $^6$Li atoms for evaporative cooling to degeneracy in a $\cdt$ can be collected in a MOT directly from an effusive atomic source and (2) the vacuum limited lifetime imposed by this atom source can be exceedingly long when the oven is properly degassed.

\section{Apparatus}
\subsection{Vacuum chamber}

The MOT trapping region is centered within an optically polished quartz cell bonded on both ends to glass to metal transitions (cell manufactured by Hellma) with the atomic source introduced on one end and supported inside the cell by a 2.75 inch conflat electrical feedthrough.  The other end of the cell is connected through a stainless steel bellows and 6 inch conflat cross to a 20 L~s$^{-1}$ Varian StarCell ion pump and a SAES CapaciTorr NEG pump.  The outer dimensions of the quartz cell are 30~mm $\times$ 30~mm $\times$ 100~mm with a 5~mm wall thickness.
\begin{figure}[ht]
  \begin{center}
    \includegraphics[width=0.49\textwidth]{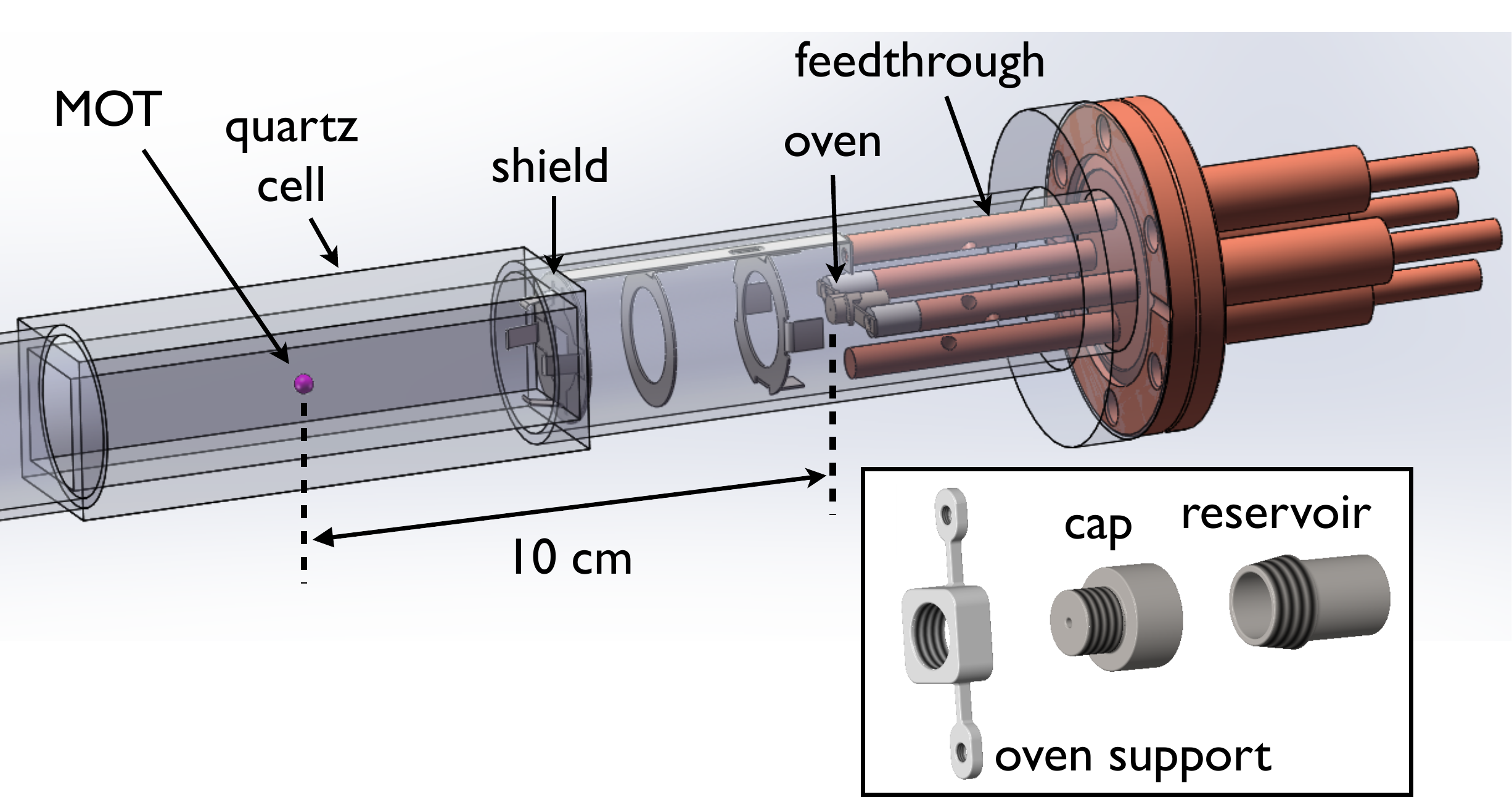}
     \caption{(Color online) Drawing of the vacuum system showing the effusive atomic source (oven), the metal shield used to prevent Li from coating the inside of the quartz cell, and the MOT position. The inset shows the 3D model of the oven.}
  \label{fig:oven}
  \end{center}
\end{figure}
The effusive lithium oven is located 10~cm from the trapping region. A small beam shield (see Fig.~\ref{fig:oven}) is situated between the oven and the trapping region in order to shield the quartz cell walls from the direct output of the effusive oven. This configuration is similar to that reported in Ref.~\cite{Anderson94} with the exception that in this work the trapping light for lithium is single frequency and is not broadened in any way to enhance the loading of the MOT.  The lithium MOT captures atoms from the low velocity tail of the effusive oven output without any additional slowing stages.

The lithium oven, made of non-magnetic stainless steel, consists of a cylindrical reservoir (15~mm long, 7~mm inner diameter, 1~mm wall thickness) enclosed by a screw cap with a 1~mm hole (Fig.~\ref{fig:oven}, inset). A supporting element is attached to the cap and is made of a non-magnetic alloy of nickel and chromium (80\% nickel and 20\% chromium). It serves as a mechanical support and electrical contact between the feedthru electrodes.  Due to the thin profile of the leads (1~mm), it provides ohmic heating for the atomic reservoir when current is applied. Both the cap and the reservoir have tapered threads commonly used to connect pipes. When properly assembled, this tapered thread ensures that hot lithium leaves only through the hole in the cap. In our previous design the connection between the pipe and the reservoir was not leak-tight and some lithium vapor was found to escape  from the sides of the oven \cite{Ladouceur:09}.

Key to achieving a good vacuum with a hot atomic oven less than 10~cm from the MOT was the careful preparation of the enriched (95\% $\LiS$) sample and systematic degassing of the oven. A nickel wire mesh was rolled onto a 4~mm rod to create a cylindrical roll 10~mm long and with an outer diameter of 7~mm and was then put inside the reservoir. The assembled oven was cleaned with acetone and baked in air to remove contaminants from the machining process. Lithium is highly reactive with both organic and inorganic reactants (including nitrogen and hydrogen); therefore, the initial sample of lithium was cleaned with petroleum ether to remove residual oil from the surface in an argon filled glove-box. A clean razor blade was used to remove the black exterior from all sides of the lithium sample.  The lithium was then loaded into the oven inside the glove box to avoid contamination of lithium by atmospheric gases. To maximize the filling of the oven and to simplify the process, the lithium sample was put in a home-built press that produced 10~mm long cylindrical extrusion with a 4~mm diameter. The extrusion was put into the center of the wire mesh roll and covered on the top with an additional round piece of nickel mesh. When heated, the melted lithium incorporates into the nickel mesh which acts as a wick preventing the molten lithium from draining out of the hot oven.

\begin{figure}[ht]
  \begin{center}
    \includegraphics[width=0.49\textwidth]{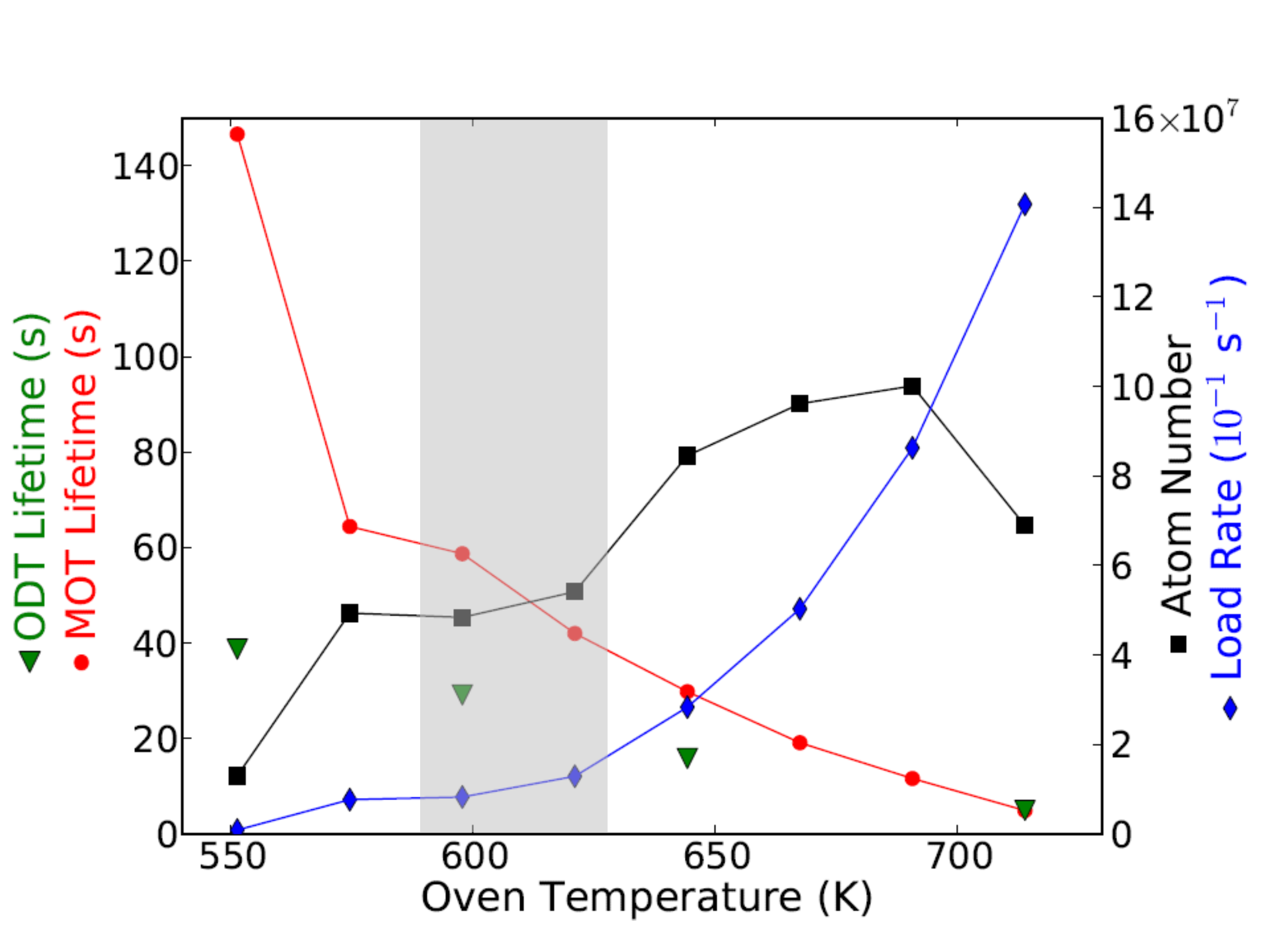}
\caption{(color online) The atom loading rate (diamonds), the MOT lifetime (circles), the ODT lifetime (triangles), and the steady state atom number (squares) of the MOT are shown as a function of the oven temperature with parameter values as defined in table~\ref{table:OptimalSettings}.  We typically operate the source in the shaded region (between 590 and 625~K) where a good compromise is found between a sufficiently large MOT atom number and a long ODT trap lifetime.  The trend lines are only a guide to the eye.}
\label{fig:OvenCurrentData}
  \end{center}
\end{figure}

The full lithium oven was degassed by heating it to 200$^{\circ}$C within an auxillary preparation chamber such that the total pressure from outgassing never exceeded $10^{-6}$~Torr. The treatment removed the majority of the trapped nitrogen, oxygen, and hydrogen. These were the primary outgassed contaminants as identified by a residual gas analyser (RGA). The oven was heated until the pressure due to outgassing at a current of 10~A (corresponding to an oven temperature of 640~K) fell below the base pressure of the preparation chamber ($5\times 10^{-8}$~Torr; 10 ~l~$\mathrm{s^{-1}}$ pumping speed). After degassing, the preparation chamber was flushed with argon (the use of dry nitrogen was avoided since it was found to contaminate the lithium) and the oven was quickly moved to the experimental vacuum chamber where a 6-day bakeout at 200$^{\circ}$C was performed. After the preparation stage about 67~mg of lithium is available. The loading performance of the lithium MOT located 10~cm from the oven is shown in Fig.~\ref{fig:OvenCurrentData}.
\subsection{Laser system}

The ``pump" light (near the $2s_{1/2}, \, F=3/2 \rightarrow 2p_{3/2}, \, F=5/2$ transition frequency) and the ``repump" light (near the $2s_{1/2}, \, F=1/2 \rightarrow 2p_{3/2}, \, F=3/2$ transition frequency) for lithium laser cooling originates from two tapered amplifiers (TA) from Sacher Lasertechnik. Each of them is seeded with 14~mW of light from a Mitsubishi ML101J27 laser diode~\footnote{These diodes emit light at 660~nm when at a temperature of 25$^{\circ}$C.  By heating them to 72$^{\circ}$C, their free running wavelength is between 670 and 671~nm.  At this temperature, the output power is significantly reduced and we observe an output of 60~mW at 314~mA.  This current is well above the damage threshold current at room temperature but does not appear to damage the diodes when they are above 68$^{\circ}$C.}, which in turn is injection locked with 2.6~mW of seed light from the master laser (Toptica DL Pro) locked 50~MHz above the pump transition. To reach required frequencies for the atom trapping the light seeding the pump TA is up-shifted by 108~MHz and then the output of the TA is down-shifted with an acousto-optic modulator (AOM) in a double pass configuration. The output of the repump TA is up-shifted with a double pass AOM. Finally, the pump and the repump light are each coupled into a separate, 50~m long, single mode, polarization maintaining fiber and sent to the experimental table. 80~mW of pump and 60~mW of repump light is available for experiments.

The hyperfine level spacings in the excited state in lithium are so small that the pump light, which is typically tuned to the red of the transition by several linewidths, excites with similar rates all three excited states, and the upper, $F=3/2$, ground state is rapidly depleted by optical pumping.  In this case, the repump light must have a similar scattering rate as the pump light and it therefore contributes to cooling and exerts a force on atoms in the MOT comparable to that exerted by the pump light.  Consequently, both the pump and repump light must be introduced into the trap along all three directions with similar intensities for the proper operation of the lithium MOT.

\begin{figure}[ht]
  \begin{center}
    \includegraphics[width=0.49\textwidth]{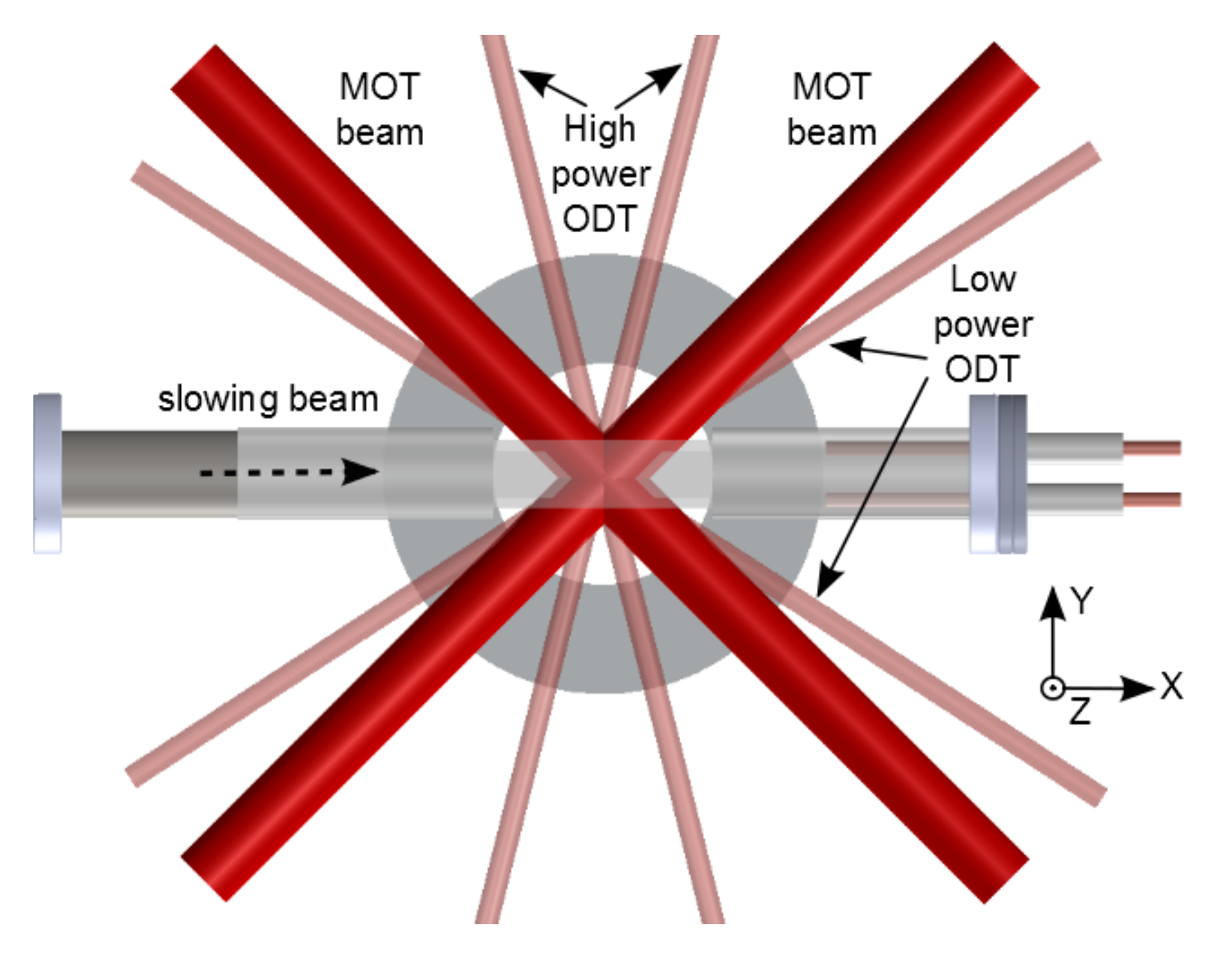}
     \caption{(Color online) Geometry and position of the MOT beams, the high and low power ODT beams and the lower magnetic coil relative to the glass cell. Not shown is the third MOT beam along the $z$-axis, the top magnetic coil and the small ``gradient" coil placed inside and concentric to it. The atomic source is attached to the feedthrough on the right side of the image (see Fig.~\ref{fig:oven}). For reference, gravity points in the $-\hat{z}$ direction and the $x$--$y$ plane is parallel to the optical table surface.}
  \label{fig:geometry}
  \end{center}
\end{figure}

The pump and repump light for trapping are combined into a single beam and, by way of several polarization beam splitter cubes, is split into four beams. Three beams are then expanded to a $1/e^2$ diameter of 2.5~cm, and introduced into the MOT cell along three mutually orthogonal axes in a retro-reflection configuration, as shown in Fig.~\ref{fig:geometry}. This configuration was chosen over a 6 independent beam configuration in order to maximize the optical power available for trapping.  The fourth (slowing) beam is sent counter-propagating to the atomic beam, and enters the cell through a viewport at the opposite end of the chamber from the oven. Because the oven output is not an intense, collimated atomic beam and because the distance from the source to the viewport is large (on the order of 1~m), we do not observe any coating of the window by Li. This effectively creates a 7-beam MOT, where the slowing beam increases the capture velocity of the MOT along the atomic beam axis, leading to about two-fold improvement in the final atom number. The quadrupole field for the MOT is produced by a pair of coils in an anti-Helmholtz configuration and whose axis of cylindrical symmetry is aligned vertically with respect to the table (see Fig.~\ref{fig:geometry}).  These are the same coils used to produce large, homogenous magnetic fields to access the Feshbach resonance during evaporation from the ODT.
For the experiments with degenerate gases reported here, the atom number was measured by absorption imaging.  The absorption probe beam included only pump light while a counter-propagating repump beam was included to provide hyperfine repumping with a radiation pressure opposing the probe beam to limit the acceleration of the atoms during imaging.

At magnetic fields within 200~G of the wide (300~G) resonance at $\frpos$ \cite{PhysRevLett.110.135301}, the Feshbach molecules of lithium are sufficiently loosely bound that it is possible to image them with standard absorption imaging \cite{PhysRevLett.91.250401}. The binding energy is so small that it introduces a negligible frequency shift from the atomic transition, allowing the molecule's atomic constituents to be imaged separately.  When free, these constituent atoms are in the two lowest magnetic sublevels whose energy at a magnetic field of B=800~G differs by on the order of 78~MHz. At large magnetic fields, the atomic hyperfine coupling is greatly suppressed and the optical transitions used for imaging these atomic states become closed and no repump light is required.  Only one frequency, detuned 600--1000~MHz below the cooling transition is necessary for imaging.  This light is produced by an in-house-built extended cavity diode laser (based on the Roithner RLT6720MG diode laser) that is locked with a large frequency offset to the master laser.  The output is then amplified by injection locking a Mitsubishi ML101J27 diode laser, and the output beam is frequency shifted by an AOM in a double pass configuration. The frequency offset lock allows precise tuning of the imaging light frequency while the AOM acts as a shutter and allows us to change the beam frequency to image either of the two lowest spin states.  It can also be used to apply short light pulses for spin-selective removal of atoms.  The beam is fiber coupled and combined with MOT pump light for imaging at low magnetic fields using a fibre-based combiner (Evanescent Optics Inc.). The imaging axis is perpendicular to the magnetic field, and in this arrangement it is impossible to polarize the light such that it drives only either $\sigma_+$ or $\sigma_-$ transitions. To optimize the absorption signal, the polarization of the imaging beam is set to be linear and along the magnetic field axis. This leads to the underestimation of the atom number in a particular spin state by a factor of two as the imaging light is a mixture of both left and right circular polarizations and only one of these polarizations drives the transitions of interest.

\subsection{MOT performance}

The MOT numbers were optimized at an oven current of 9.5~A (corresponding to an oven temperature of 625~K) by carrying out an exhaustive search in the pump ($\dnp$) and repump ($\dnr$) detunings for a range of axial magnetic field gradients ($\amfg$) between 10 to 70~G\,cm$^{-1}$ \footnote{The axial magnetic field gradient for our coils is 5.8~G/cm/A}.  The optimum MOT parameters are summarized in table~\ref{table:OptimalSettings}.  The highest atom number we were able to trap (at $I_{\mathrm{oven}} = 9.5~\mathrm{A}$) was $\maxli$ atoms at $\amfg = 35$~G cm$^{-1}$ and using a detuning from resonance of $\dnp=-8.5 \, \GLi$ and $\dnr=-5.1 \, \GLi$ for the pump and repump light, respectively.  Note that the saturation intensities given in table~\ref{table:OptimalSettings} are calculated for the case of an isotropic pump field with equal components in all three possible polarizations \cite{Steck08}.  In addition, the lithium interaction was assumed to include all of the allowed $2^2P_{3/2}$ excited state levels due to our large detuning relative to the excited state hyperfine splittings.

\begin{table}[h]
\centering
\caption[c]{Optimal parameter settings for the MOT at an oven current of 9.5~A (oven temperature of 625~K): Wavelength and width of the D2 line, saturation intensities (assuming isotropic light polarization), pump and repump intensities along all three principle axes, relative detunings of the pump and repump light ( $\dnp$ and  $\dnr$), axial magnetic field gradient, initial load rate, decay rate, and steady state atom number.}
\label{table:OptimalSettings}
\begin{tabular}{lc}
  \hline
  \hline
  \ & $^6$Li \\
  \hline
  $\lambda_{\mathrm{D}_2,\mathrm{vac}}$ (nm) & 670.977 \\
  $\Gamma / 2 \pi$ (MHz) & 5.87 \\
  $I_{\mathrm{sat}}$ (mW~cm$^{-2}$) & 3.8 \cite{Steck08} \\
  $\Ip/I_{\mathrm{sat}}$ & 21 \\
  $\Ir/I_{\mathrm{sat}}$ & 16  \\
  $\dnp$ ($\Gamma$) & -8.5  \\
  $\dnr$ ($\Gamma$) & -5.1  \\
  $\amfg$ (G~cm$^{-1}$) & 35  \\
  $R$ & $\maxlir$ \\
  $\Nfty$ & $\maxli$ \\
  \hline   
  \hline
\end{tabular}
\end{table}

Fig.~\ref{fig:OvenCurrentData} shows the behavior of this oven loaded lithium MOT at a field gradient of 35~G\,cm$^{-1}$.  The loading rate ($R$), the lifetime or inverse loss rate ($\gl^{-1}$) of both the MOT and ODT, and the steady state atom number ($\NLis$) in the MOT are shown as a function of the oven temperature.  These parameters were determined by fitting the loading curve of the MOT to the solution, $N(t) = \NLis (1 - e^{-\gl t})$ of the differential equation $\dot{N} = R - \gl N$.  The steady state number is then the product of the loading and inverse loss rates, $\NLis = R/\gl$.

This model does not include a two-body loss term to model particle loss due to light assisted collisions between cold atoms within the MOT, and therefore the loss term we report is an overestimate of the MOT losses due only to collisions with the residual background vapor or atoms from the hot atomic beam.  In addition, because the Li--Li$^*$ collision cross section (with one atom in the excited state) is much larger than for ground state collisions, the loss rate for the excited state atoms in the MOT will be significantly larger due to collisions with fast moving Li atoms in the atomic beam.  These two effects make the MOT lifetime only an estimate of the expected lifetime for the ensemble in the $\cdt$.

For comparison, the inverse loss rate (i.e.~lifetime) of a 50~$\mu$K deep $\cdt$ is also provided for a few of the oven current settings.  Since the MOT is significantly deeper than the $\cdt$, the MOT trap loss rate due to background vapor collisions is expected to be smaller \cite{PhysRevA.80.022712,PhysRevA.84.022708}.  In addition, residual losses from the $\cdt$ will occur due to spontaneous emission and evaporation losses.  Thus the lifetime in a shallow $\cdt$ will always be an overestimate of the total background collision rate.

At low oven current settings where the loss rate is dominated by collisions with atoms or molecules other than Li atoms, the MOT lifetime is, as expected, much larger than the lifetime in the $\cdt$.  Whereas at high oven current settings where the loss rate is primarily determined by the collisions with fast moving atoms in the Li atomic beam, the MOT lifetime is similar to the $\cdt$ lifetime.  The longest inverse loss rate measured for the $\cdt$ was on the order of 40~s (at an oven current of 8~A and a temperature of 550~K) corresponding to a background vapor pressure of approximately $10^{-10}$~Torr \cite{Metcalf,PhysRevA.80.022712,PhysRevA.84.022708}.  As the temperature of the oven increases so does the captured flux and the loss rate.  At a current of $I_{\mathrm{oven}} = 11~\mathrm{A}$ (corresponding to an oven temperature of 680~K), the steady state atom number is maximized. There exists a clear tradeoff between the inverse loss rate of the lithium MOT and the steady state atom number, and by running the oven at 625~K (with $I_{\mathrm{oven}} = 9.5~\mathrm{A}$) instead of 680~K, the inverse loss rate can be increased by more than a factor of four with only a factor of two reduction in the trapped number.  We note that the MOT performance is similar to that previously reported in Ref.~\cite{Ladouceur:09}.  The primary difference is that here the oven aperture is 30\% larger and a portion of the MOT light is sent counter propagating to the oven flux.  This additional slowing beam enhances the loading rate by a factor of two.

\subsection{Optical Dipole Trapping}

In this experiment, a high power but broad linewidth laser creates an optical dipole trap (200 W total power) that is loaded from the MOT.  A large power and small beam waist is necessary to achieve a 3.5~mK ODT depth required to capture the atoms from the MOT.  The atoms are then transferred to a shallow, narrow linewidth, low power $\cdt$ (18~W total power). This is done because an eventual goal is to confine the atoms in a 1D optical lattice formed by the narrow linewidth ODT. However, we note that this procedure is not optimal in terms of the final atom number at degeneracy since the transfer between the ODTs is only 25\% efficient. The geometry of the high and low power ODTs relative to the glass cell is shown in Fig.~\ref{fig:geometry}.

In order to transfer the atoms into the $\cdt$, we compress and cool the atoms in the MOT by increasing the axial magnetic field gradient from $35$ to $64$~G cm$^{-1}$, lowering the intensity and shifting the frequency of both the pump and repump light to 10~MHz below resonance.  During this compression and cooling phase, a $\cdt$ of 200 W total power is turned on and, in less than 10~ms,  up to 10\% of the $^{6}$Li atoms are transferred into the $\cdt$.  Fine adjustments of the MOT position to optimize its overlap with the $\cdt$ are essential to achieving efficient transfer and are made by adding a small (5 to 10~G) homogeneous magnetic field that shifts the magnetic quadrupole center during the compression phase.  This field is generated by three pairs of large (30~cm diameter) ``compensation" coils for independent control of the field along each of the three orthogonal spatial axes.

We observe extremely rapid trap losses (with trap lifetimes on the order of a few ms) due to light assisted collisions and hyperfine relaxation, and we therefore optically pump to the lower hyperfine state ($F=1/2$) during the transfer by extinguishing the repump light 400~$\mu$s before the pump light.  This procedure produces an almost equal population of the two sub-levels of the lower hyperfine state : $\one \equiv |F=1/2, m_F=1/2\rangle$ and $\two \equiv |F=1/2, m_F=-1/2\rangle$.  The light for the $\cdt$ is derived from a multi-longitudinal mode, 100~W fiber laser (SPI Lasers, SP-100C-0013) operating at 1090 nm with a spectral width exceeding 1 nm.  The $\cdt$ is comprised of two nearly co-propagating beams crossing at an angle of 14$^{\circ}$.  Each beam has a maximum power of 100~W (for a total power of 200 W achieved by recycling the first beam) and is focused to a waist ($1/e^2$ intensity radius) of $42$~$\mu$m and $49$~$\mu$m.  After the MOT light is extinguished, the $\cdt$ beam power is ramped down linearly in time to 100 W total (50 W per beam) in 100~ms while applying a homogenous magnetic field of 300~G produced by the same coils used to generate the quadrupole magnetic field for the MOT. The output power of this SPI laser is varied by an analog input controlling the pump diode current of the post amplifier stage.  Rapid thermalization occurs because of the large collision rate between the $\one$ and $\two$ states at 300~G \cite{PhysRevLett.110.135301}.  At the end of this forced evaporation stage, there are approximately $10^6$ atoms remaining at a temperature of $200$~$\mu$K (verified by a time-of-flight expansion measurement).

We observe thermal lensing effects which produce beam aberrations of our high power $\cdt$.  In particular, we observe the position of the beam waists of the $\cdt$ arms moving at approximately 780~$\mu$m~s$^{-1}$ when the high power $\cdt$ is initially turned on with 100~W per beam \footnote{The beam focus is observed to move 395~$\mu$m in the first 500~ms and 280~$\mu$m in the next 500~ms due to thermal lensing}.  These thermal effects were worse with a vacuum cell made from borosilicate glass.  For this reason, we are now using a quartz cell.  Not only does quartz have a higher transmission in the near IR (e.g.~at 1090 nm), but the thermal expansion coefficient of quartz is ten times smaller than that of borosilicate glass.

At the end of the pre-evaporation stage the atoms are transferred into a lower-power $\cdt$ whose intersection is aligned to the crossing of the high-power $\cdt$.  The light for this second $\cdt$ is generated by a single frequency, narrow-linewidth ($<10$ kHz), 20 W fiber laser operating at 1064~nm (IPG Photonics YLR-20-1064-LP-SF).  This transfer is done to avoid ensemble heating observed in the SPI laser $\cdt$ and allows further forced evaporative cooling to much lower powers and thus ensemble temperatures.  In addition, as mentioned already, this narrow line-width laser trap can be used to realize an optical lattice.  The disadvantage of this step is that the transfer is not yet well optimized and we only retain 25\% of the atoms. Further optimization of this transfer stage by aligning the trapping beams to be co-linear (in order to increase the spatial overlap of the trapping potentials due to each laser) is expected to produce significant improvement in the final atom number at the transition to quantum degeneracy.

The IPG $\cdt$ is comprised of two beams in the horizontal plane and crossing at an angle of 60$^{\circ}$ and with a total power of 18~W (9~W per beam).  The beams are focused to a waist ($1/e^2$ intensity radius) of $25$~$\mu$m and $36$~$\mu$m.  The polarization of the two beams (which have the same frequency) is chosen to lie in the horizontal plane to minimize reflections at the cell (the cell is uncoated but the beams enter the cell close to Brewster's angle).  Unexpected heating and loss was observed when the polarization of the second beam was fixed perpendicular to the first (i.e.~perpendicular to the horizontal plane and parallel to the magnetic field axis).  The IPG $\cdt$ beam power is controlled by a water cooled AOM (Intraaction ATD-1153DA6M). The diffraction efficiency of the AOM is 60\%. Due to the experimental constraints the polarization of the IPG $\cdt$ is perpendicular to the polarization that, according to the manufacturer of the AOM, would lead to the diffraction efficiency of 75\%.  After the transfer of atoms into the IPG $\cdt$ further evaporation is done at a field specific to a given experiment (below $\frpos$ to create molecules and above $\frpos$ to create a strongly interacting Fermi gas).  It is worth noting here that the power of the IPG laser is not actively stabilized.

\section{Realization of Quantum Degenerate Gases}

Over the past decade, quantum degenerate Fermi gases of either $^6$Li or $^{40}$K atoms have enabled the experimental study of the so-called Bardeen-Cooper-Schrieffer--Bose-Einstein condensation crossover physics (see \cite{Regal20061} and references therein).  In the following, we present data obtained with this apparatus showing evidence of both Bose Einstein condensation of molecules and pairing in a strongly-interacting, degenerate Fermi gas.  The goal here is not to quantitatively interpret these data, but rather to present representative data obtained with this apparatus showing signatures of these phenomena studied in detail previously.

\subsection{Evaporative cooling to degeneracy}

\begin{figure}[ht]
  \begin{center}
    \includegraphics[width=0.49\textwidth]{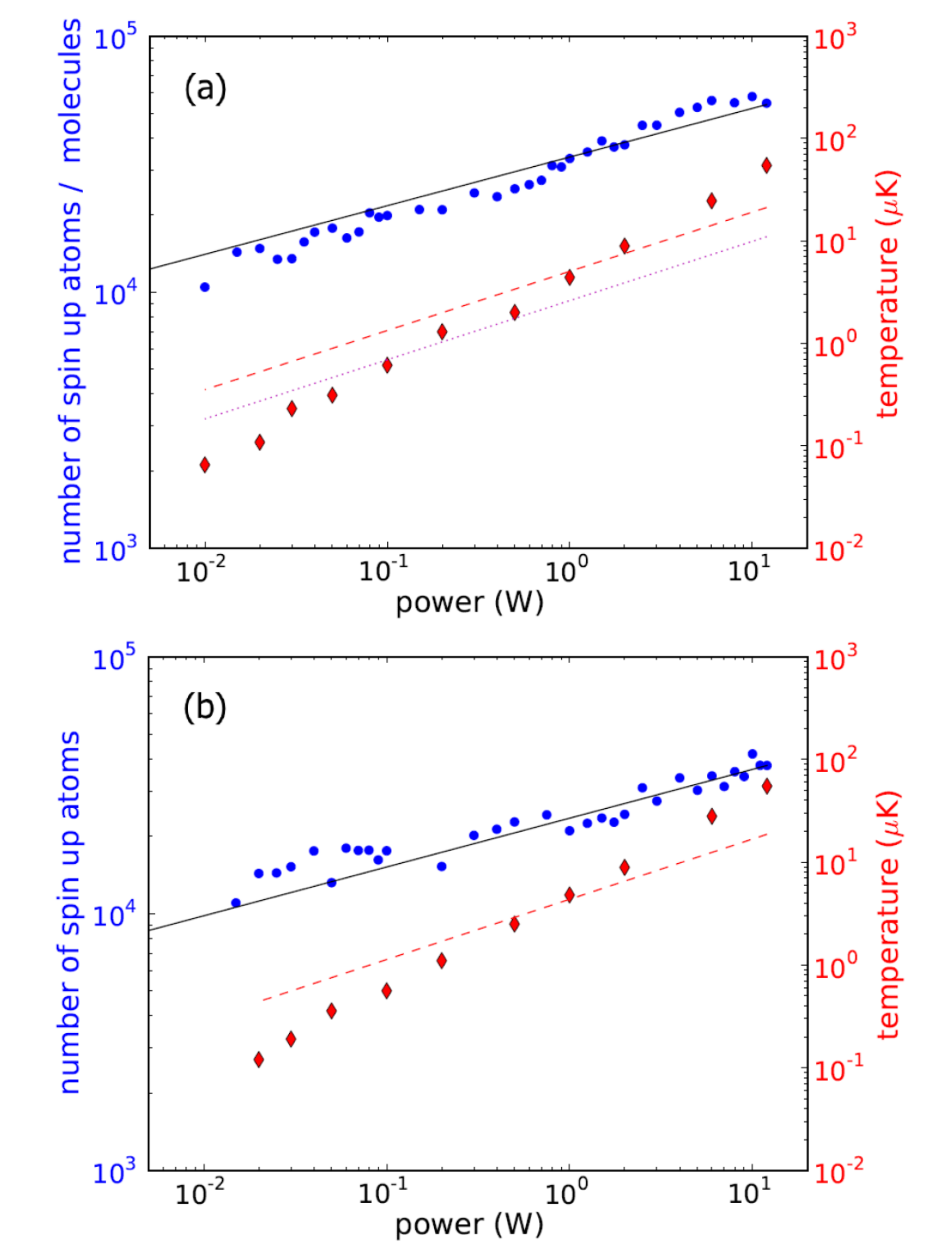}
     \caption{(Color online) Ensemble temperature (diamonds) and particle number (circles) as a function of the $\cdt$
     laser power (i.e.~trap depth) in the final forced evaporation stage at
     (a) B=800~G (below the FR) and (b) B=900~G (above the FR).  The number of ``spin-up" atoms
     (atoms in the $|1\rangle$ state) is detected at these high magnetic fields by absorption imaging.  
     For the evaporation below the FR, when the temperature is well below the binding energy all of the
     atoms form FR molecules and the number detected in the $|1\rangle$ state corresponds to the total FR molecule number.
    The solid line indicates the expected particle number for an efficient evaporative cooling.
    The dashed line indicates the Fermi temperature for the $|1\rangle$ component at each power computed from the particle number and the trap frequencies.
     In (a) the dotted line shows the critical temperature for Bose Einstein condensation for the molecules.
      }
  \label{fig:evap}
  \end{center}
\end{figure}

The trap depth is lowered using a series of linear ramps of different duration, for a total evaporation time of 4~s. We use an additional ``gradient" coil, concentric with the top main magnetic coil, to add a magnetic field gradient to both compensate for the residual magnetic field gradient of our main magnetic coil pair and to provide a magnetic force on the atoms equal and opposite to the gravitational force. This additional coil is essential for the production of quantum degenerate samples as in this case the evaporation must be continued to the very lowest optical trap depths where gravitational sagging significantly compromises the confinement.

Figure~\ref{fig:evap} shows the result of evaporation both below and above the Feshbach resonance center at $\frpos$.  Both the ensemble temperature and particle number are shown as a function of the trap depth during the final forced evaporation stage.  The solid line indicates the expected particle number for an efficient evaporative cooling where the ensemble temperature remains a factor of ten below the trap depth ($\eta= \ut/\kb T=10$).  For a discussion of the scaling laws for evaporative cooling in time-dependent optical traps, see Ref.~\cite{PhysRevA.64.051403}.

\subsection{Molecule production and condensation}

\begin{figure}[ht]
  \begin{center}
    \includegraphics[width=0.49\textwidth]{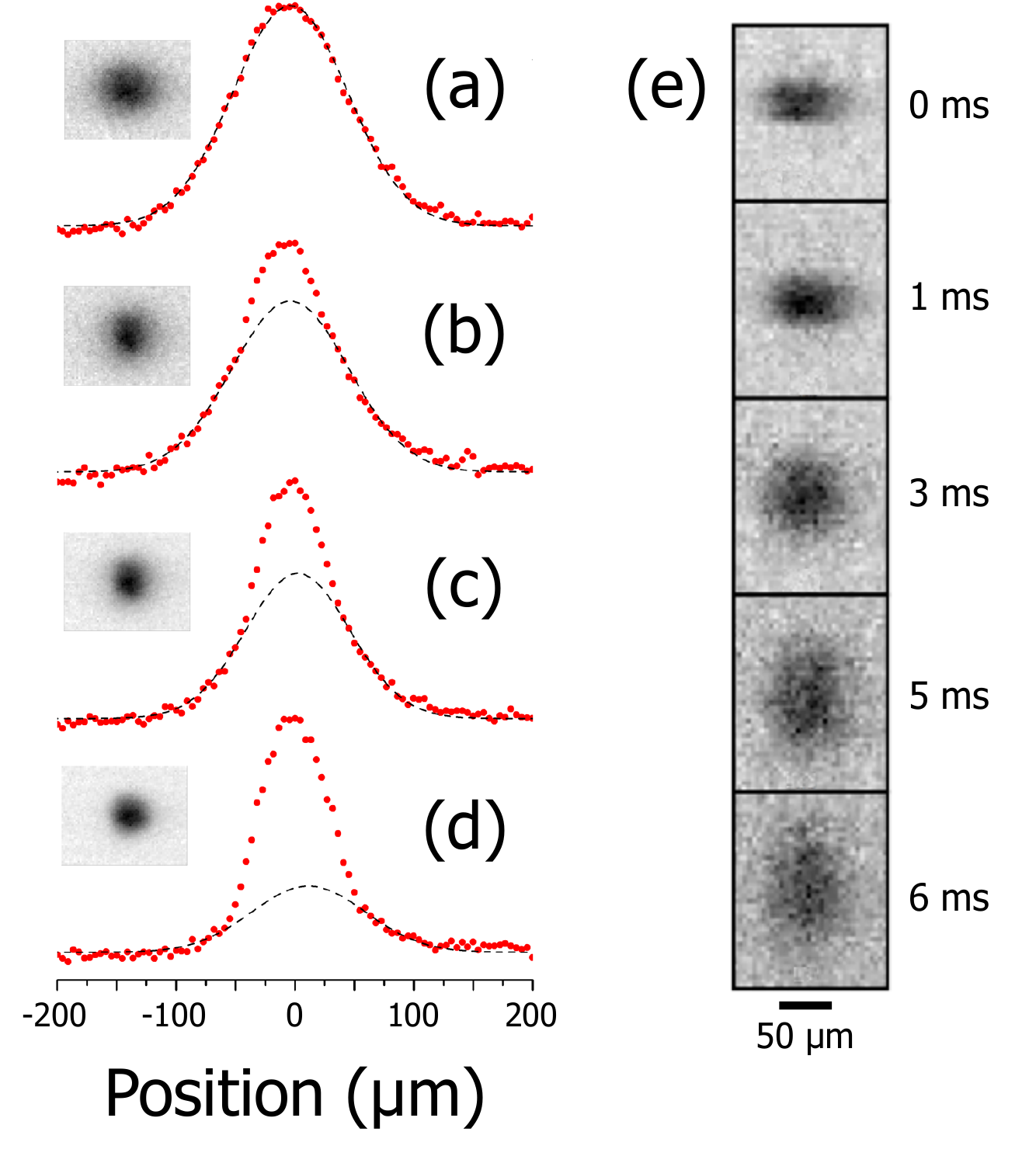}
     \caption{(Color online) Absorption images
     and resulting horizontal density profiles of $^6$Li$_2$ Feshbach molecules after a 3~ms time of flight expansion at a magnetic field of $B=\btof$
     for different ensemble temperatures, (a) $T=710$~nK ($T/\tc \sim 1.2$), (b) $T=230$~nK ($T/\tc \sim 0.9$), (c) $T=110$~nK ($T/\tc \sim 0.8$), (d) $T=65$~nK ($T/\tc \sim 0.6$).  The wings of the profiles are
     fit to a Gaussian function and the bimodal character of the density profile is evident for temperatures well below the 
     critical temperature.  For the coldest molecular cloud (d) we show in (e) the absorption images after different
     free expansion times.  The in-situ shape of the molecular cloud (image taken at 0 ms) has the anisotropic shape of the $\cdt$, and the cloud
     anisotropy reverses during the time-of-flight expansion.
      }
  \label{fig:mBEC}
  \end{center}
\end{figure}

For evaporation at $B=800$~G, we observe the formation of FR molecules when the temperature nears the molecule binding energy (250~nK), and at the end of the evaporation ramp, we observe $2 \times 10^4$ molecules at a temperature of 70~nK.  To image the molecular cloud, we reduce the magnetic field from 800 to $\btof$ and release the ensemble from the dipole trap.  After some fixed time-of-flight (TOF) expansion, we then take an absorption image of the molecular cloud.  As mentioned before, the spin-up (or spin-down) component of this very weakly bound molecule can be imaged as if it were a free atom since the binding energy is far below the excited state width.

At the end of the evaporation ramp at $B=800$~G the molecular gas is strongly interacting and deep in the hydrodynamic regime.  In particular, the molecule-molecule $s$-wave scattering length is predicted to be on the same order as the atom-atom scattering length ($\amol=0.6a$ \cite{PhysRevLett.93.090404}), and the atom-atom scattering length at B=800~G is $a>5000 \aB$ \cite{PhysRevA.66.041401}.  In order to reduce the interaction strength during the TOF expansion and thus increase the visibility of the characteristic bimodal distribution of the mBEC, the TOF is performed at $\btof$ \cite{PhysRevLett.92.120401}.

Figure~\ref{fig:mBEC} shows a series of absorption images and resulting horizontal density profiles of these Feshbach molecules after a 3~ms TOF expansion at a magnetic field of $B=\btof$ for different final ensemble temperatures.  The wings of the profiles are fit to a Gaussian function and the bimodal character of the density profile is evident for temperatures below the critical temperature for Bose Einstein condensation.  For the coldest molecular cloud, we show in Fig.~\ref{fig:mBEC} the absorption images after different free expansion times.  The in-situ shape of the molecular cloud (image taken at 0 ms) has the anisotropic shape of the $\cdt$, and the cloud anisotropy reverses during the time-of-flight expansion.  This anisotropy reversal is expected for a BEC of non-interacting particles due to the larger confinement frequency along the vertical direction in these images.  However, even at the magnetic field of $B=\btof$, the molecules are still strongly interacting, and the inversion of the aspect ratio of the molecular cloud can also be the result of hydrodynamics \cite{O'Hara13122002,PhysRevLett.106.115304}.

\begin{figure}[ht]
  \begin{center}
    \includegraphics[width=0.49\textwidth]{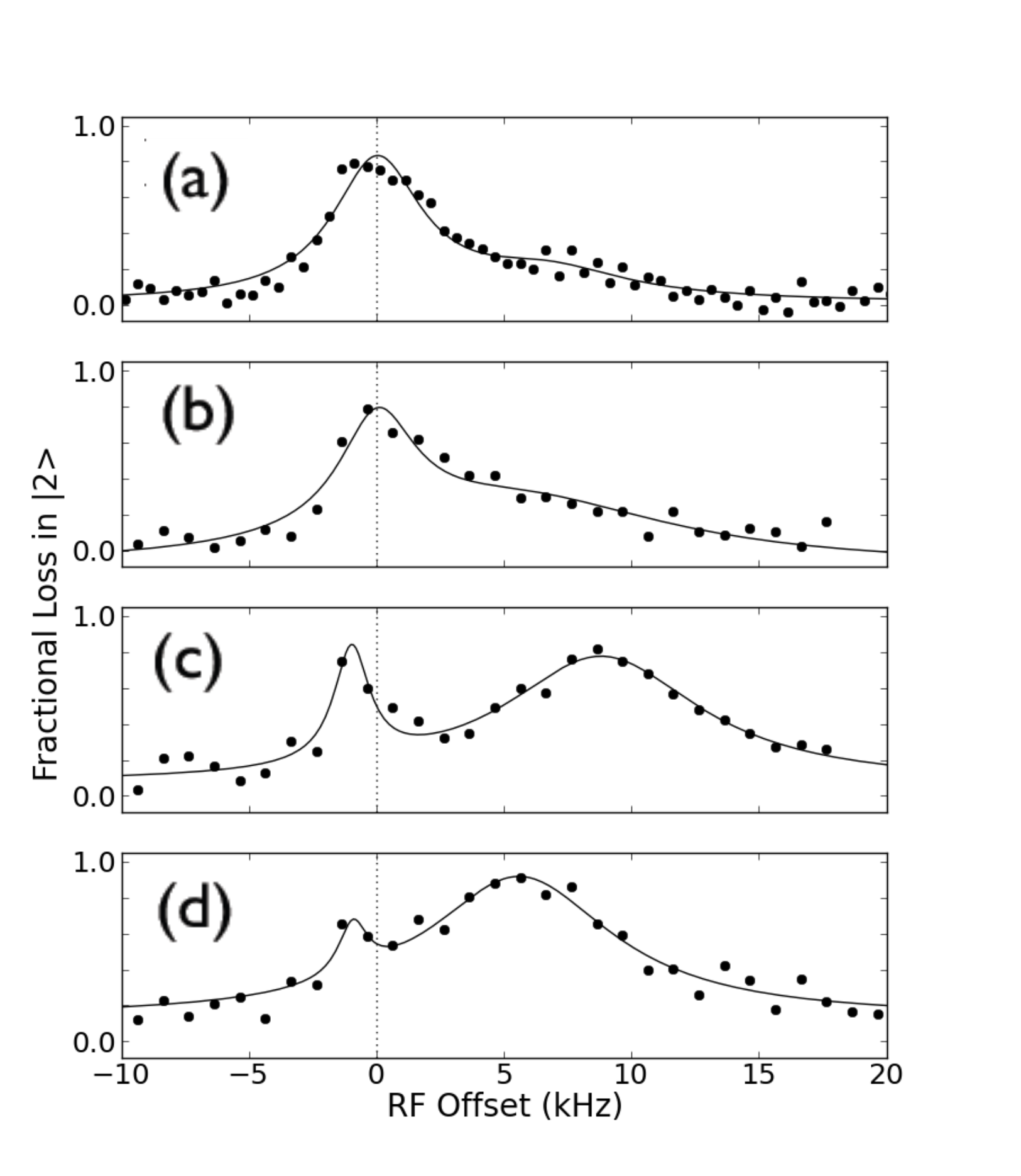}
     \caption{Radio frequency (rf) spectroscopy of a strongly interacting two component Fermi gas at $B=839.2$~G. The number of atoms lost from the $|2\rangle$ state is plotted after the application of an rf field as a function of its frequency offset from the unpaired atom resonance.  In (a) and (b) where the temperature is above the pair dissociation temperature, $\td$, the rf-spectrum shows a single resonance. This corresponds to the energy required to flip the spin of an unpaired atom.  In (c) and (d), the ensemble temperature is $T<\td$, and a second maximum appears at higher energy due to pair formation.  Here the temperature is slightly above the critical temperature for pair condensation, and these so-called pre-formed pairs exist in what is known as the "pseudogap" region \cite{Chen20051,PhysRevA.84.011608}.  In (d) where the temperature is lower than in (c), the ratio of pairs to free atoms is larger as expected; however, the gap energy is less because the Fermi temperature in (d) is smaller than in (c). The lines show a fit of the data to two lorentzians and are meant only as a guide to the eye.
     }
  \label{fig:pairing_gap}
  \end{center}
\end{figure}

\subsection{Fermionic pairing measurements}

One of the first experimental signatures of fermionic pairing above the FR where the atom-atom interactions are attractive was the observation of the pairing gap in a strongly interacting Fermi gas of $^6$Li atoms by radio frequency spectroscopy \cite{Chin20082004}.  While the correct interpretation of these spectra requires a full accounting of both the final-state and trap effects to understand the contributions from the pairing-gap, pseudogap, and no-gap phases, nevertheless, it is a simple measurement that can be intuitively understood and that provides a clear signature of pair formation (both so-called ``pre-formed" and condensed pairs) \cite{PhysRevA.84.011608}.

Here we perform the final evaporation stage at $B=839.2$~G, just above the $\fr$ to different final trap depths producing different final ensemble temperatures.  At the end of this evaporation, we apply radio frequency (rf) radiation for 1 second that transfers atoms in state $|2\rangle$ to the unoccupied $|3\rangle$ state.  We monitor the loss of atoms from state $|2\rangle$ by state-selective absorption imaging, and in Fig.~\ref{fig:pairing_gap} we plot the loss as a function of the rf radiation frequency detuning $\delta$ from the free atom resonance at 81.34 MHz (corresponding to the energy splitting between the $|2\rangle$ and $|3\rangle$ states at this magnetic field including the difference in the mean field interaction energies).  We note that when the loss of atoms from state $|1\rangle$ is monitored instead of the loss from state $|2\rangle$ the spectrum is the same.  This is likely due to the loss of atoms in state $|1\rangle$ due to the collisional instability of these mixtures when $|3\rangle$ state atoms are present \cite{PhysRevLett.101.203202,PhysRevLett.103.130404}.

In Fig.~\ref{fig:pairing_gap}(a) and (b) where the temperature is above the pair dissociation temperature, $\td$, the rf-spectrum shows a single resonance peak at a frequency offset of zero.  This corresponds to the energy required to flip the spin of an unpaired atom.  In (c) and (d), the ensemble temperature is $T<\td$, and a second maximum appears at higher energy.  Unpaired $|2\rangle$ atoms undergo the transition at a zero offset whereas an additional energy due, in part to the pairing gap, must be added to flip those bound to a $|1\rangle$ state atom as a pair.  Here the temperature is slightly above the critical temperature for pair condensation, and these so-called pre-formed pairs exist in what is known as the "pseudogap" region \cite{Chen20051,PhysRevA.84.011608}.  In Fig.~\ref{fig:pairing_gap}(d) where the temperature is lower than in (c), the ratio of pairs to free atoms is larger as expected; however, the gap energy is less because the Fermi temperature in (d) is smaller than in (c).  In addition, the unpaired spin flip energy is shifted to a negative offset in (c) and (d).  This is because we produce a colder ensemble by evaporating to a lower trap depth and this changes the atomic density producing a different differential mean field interaction energy for the unpaired atom transition \cite{Gupta13062003}.

\section{Conclusions}
\label{sec:conc}

In summary, we demonstrate that the conditions for achieving a quantum degenerate gas can be met for $^6$Li with a simple oven-loaded MOT.  Our apparatus does not require a Zeeman slower or a 2D MOT nor does it require any separation or differential pumping between the effusive atom source and the science chamber.  The result is a simple, inexpensive, and compact vacuum system ideal for miniaturization.  We transfer $1.3 \times 10^6$ $^6$Li atoms into a $\cdt$ from a MOT of $2 \times 10^7$ atoms accumulated from the atomic beam flux in 20~s.  We demonstrate with this system the production of a mBEC, and we use it to observe evidence of the pairing gap in a strongly interacting two-component DFG in the BEC-BCS crossover regime.  The key results of this work are that (1) a sufficient number of $^6$Li atoms for evaporative cooling to degeneracy in a $\cdt$ can be collected in a MOT directly from an effusive atomic source and (2) the vacuum limited lifetime imposed by this atom source can be exceedingly long when properly degassed.


The authors gratefully acknowledge Florian Schreck for helpful advice and Abraham Olson for useful feedback on this manuscript.
The authors also acknowledge financial support from the Canadian Institute for Advanced Research (CIfAR), the Natural Sciences and Engineering Research Council of Canada (NSERC / CRSNG), and the Canadian Foundation for Innovation (CFI).  This work was done under the auspices of the Center for Research on Ultra-Cold Systems (CRUCS).


\end{document}